\documentclass[12pt]{article}
\usepackage{epsfig}

\newcommand{\mysection}{\setcounter{equation}{0}\section}

\def\beq{\begin{equation}}
\def\eeq{\end{equation}}
\def\beqa{\begin{eqnarray}}
\def\eeqa{\end{eqnarray}}

\newlength{\dinwidth} \newlength{\dinmargin}
\begin{document}

\begin{center}
{\Large \bf Theoretical results for top quark production}
\end{center}
\vspace{2mm}
\begin{center}
{\large Nikolaos Kidonakis}\\
\vspace{2mm}
{\it Kennesaw State University,  Physics \#1202,\\
1000 Chastain Rd., Kennesaw, GA 30144-5591, USA} \\
\end{center}

\begin{abstract}
I discuss and compare several approaches to higher-order calculations 
of top quark production. I study the relative effectiveness of the approaches 
in approximating the exact NNLO results for the top-pair total cross 
section and highlight the theoretical and numerical differences between them. 
I show that the results from my soft-gluon resummation method are nearly 
identical to the exact NNLO cross section at all LHC and Tevatron energies.
This agreement has important consequences for the validity of existing 
approximate NNLO differential distributions, for further refinements of the 
predictions, and for applications to other processes such as single-top 
production. I also compare approximate NNLO top quark transverse momentum  
and rapidity distributions with recent LHC measurements.
\end{abstract}

\mysection{Introduction}

Top quark production has been a topic of intense theoretical study for a long 
time, with NLO calculations \cite{NLO} appearing over two decades ago. 
Fixed-order calculations plus soft-gluon resummations of 
various kinds have been employed to predict the theoretical top-pair 
production cross section as well as various differential distributions. 
The theoretical formalisms that have been used for resummed calculations 
and resultant approximate NNLO results have large differences in scope 
and give a wide range of numerical values for phenomenologically interesting 
total and differential cross sections.

The recent calculation of the exact numerical NNLO total cross section 
\cite{NNLO} affords the possibility of evaluating the relative success 
of the various resummation/approximate NNLO approaches 
\cite{NK,ALLMUW,AFNPY,BFKS,CCMMN} and 
of drawing implications for differential quantities as well as for other 
related processes. This is the topic of this contribution.

\mysection{A comparison of NNLO results for the top-pair cross section}

\begin{table}[h]
\begin{center}
  \begin{tabular}{c|c}
cross section & Soft limit \\ \hline 
1PI \hspace{2mm} $d\sigma/dp_T dy$ & $s_4
    = s+t_1+u_1 \rightarrow 0$ \\ 
PIM \hspace{2mm} $d\sigma/dM_{t\bar t}d\theta$ & $1-z = 1-M_{t\bar t}^2/s \rightarrow 
    0$ \\ 
total \hspace{3mm} $\sigma$ & $\beta = \sqrt{1-4m_t^2/s} \rightarrow 0$ \\
\hline
\end{tabular}
\end{center}
\caption{
The double-differential cross sections, in single-particle-inclusive (1PI) and 
pair-invariant-mass (PIM) kinematics, and the total cross section for which 
soft-gluon resummation have been developed. The variables that vanish in 
the soft limit for each case are indicated.}
\end{table}

A lot of work has been done in the last twenty years on soft-gluon resummations
which have culminated in NNLL accuracy for top-pair production in various 
distinct approaches \cite{NK,ALLMUW,AFNPY,BFKS} (for more details and 
references on the development of resummation see the review in 
Ref. \cite{NKBP}). The resummed expressions have 
been used as generators of approximate NNLO corrections, and cross section 
calculations for top-pair production have appeared in 
\cite{NK,ALLMUW,AFNPY,BFKS,CCMMN}.
The differences between these various resummation/NNLO approximate approaches 
include: 

$\bullet$ Double-differential cross sections \cite{NK,AFNPY} versus 
total-only cross sections \cite{ALLMUW,BFKS,CCMMN}. 
These involve different definitions of threshold, see Table 1. 

$\bullet$ Moment-space perturbative QCD (pQCD) \cite{NK,ALLMUW,CCMMN} versus 
Soft-Collinear Effective Theory (SCET) \cite{AFNPY,BFKS}.

The more general approach is double-differential which allows the 
calculation of transverse momentum and rapidity distributions, as well as the 
total cross section by integrating over $p_T$ and rapidity. This approach 
uses partonic threshold, i.e. the top quark is not necessarily produced 
at rest but can have arbitrarily large velocity.
On the other hand, the total-cross-section-only  approaches use 
absolute/production threshold (top produced at rest) and are thus a 
limit/special case of the more general partonic threshold. 
A detailed discussion of these matters can be found in 
Ref. \cite{NKBP} (see also \cite{NK12,NKckm12}).
Further differences between the formalisms  
arise from what subleading terms are included, whether damping factors 
are used, and, for differential calculations, how the partonic threshold 
relation $s+t_1+u_1=0$ is used in the plus-distribution
coefficients (again see Refs. \cite{NKBP} and \cite{NK12,NKckm12} for details).
It is very important to note that while many of these differences are 
formally ``subleading'' they can be numerically very significant. 

\begin{figure}
\vspace{2mm}
\begin{center}
\includegraphics[width=10cm]{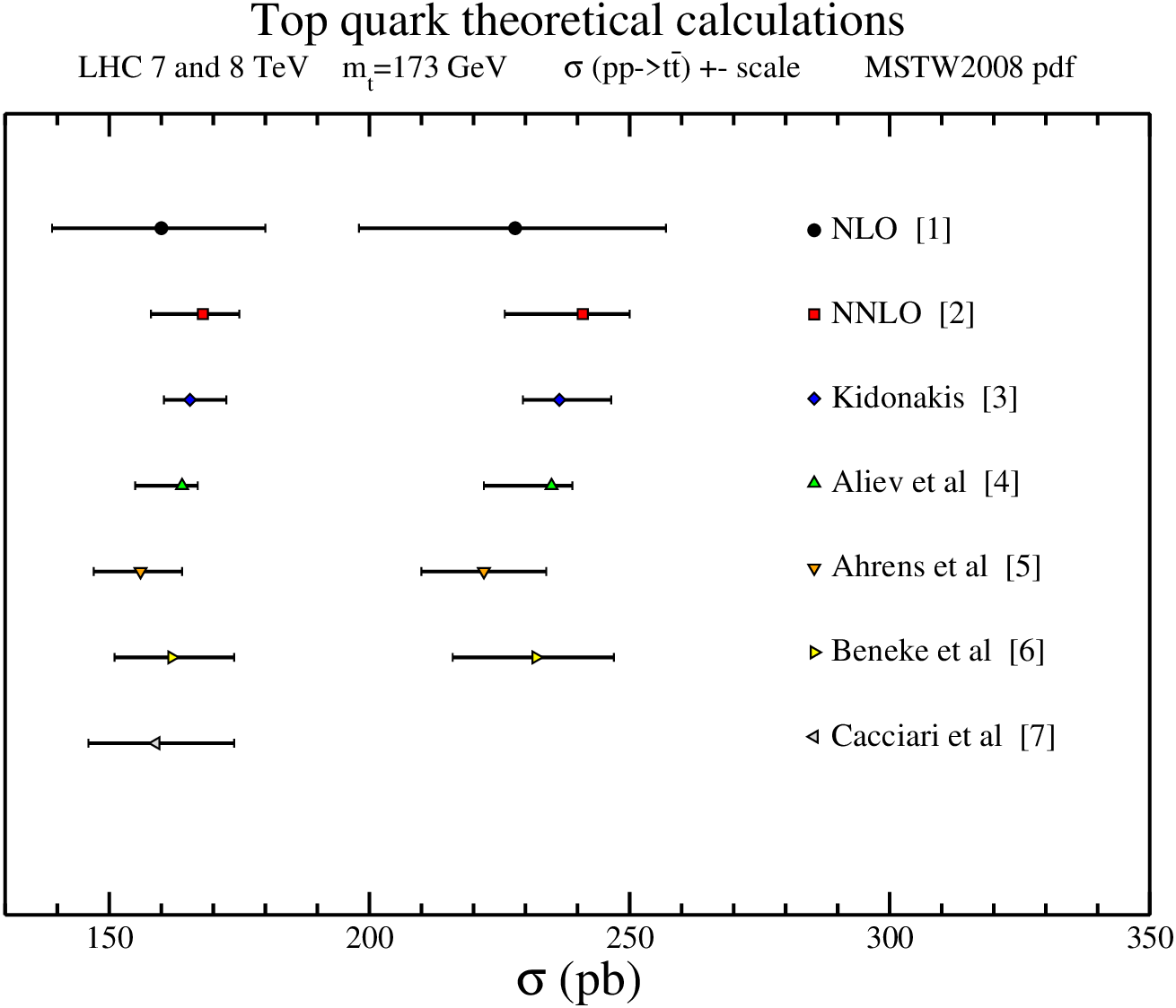}
\caption{NNLO exact \cite{NNLO} and approximate [3-7] results for the 
$t{\bar t}$ cross section at 7 TeV (bars on the left) and 8 TeV (right) 
LHC energy.}
\label{theory7lhcplot}
\end{center}
\end{figure}

\begin{figure}
\begin{center}
\includegraphics[width=80mm]{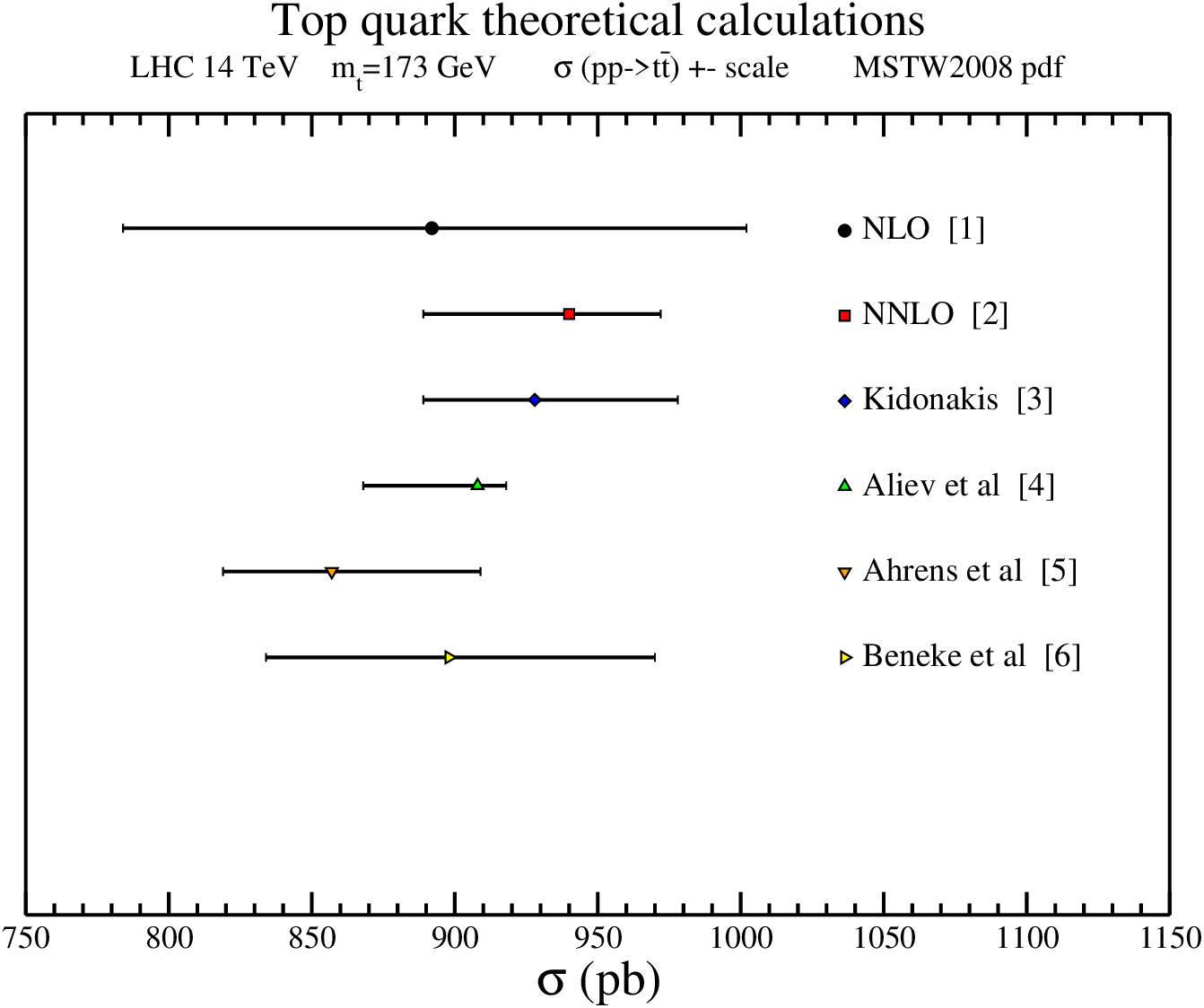}
\hspace{8mm}
\includegraphics[width=77mm]{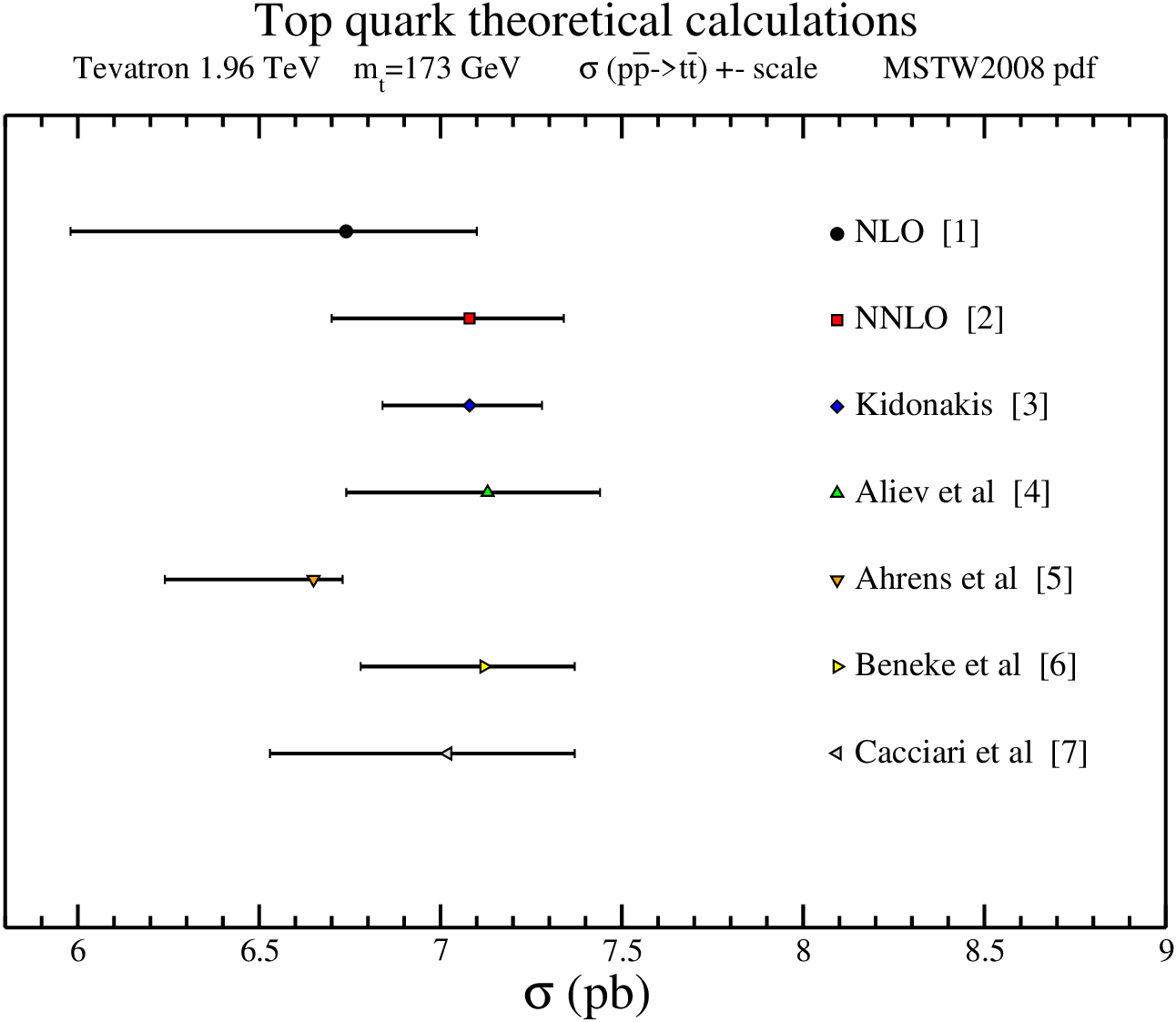}
\caption{NNLO exact \cite{NNLO} and approximate [3-7] results for the 
$t{\bar t}$ cross section at 14 TeV LHC energy (left) and at 
the Tevatron (right).}
\label{theorytev14lhcplot}
\end{center}
\end{figure}

In Fig. \ref{theory7lhcplot} we provide a comparison of various NNLO 
approximate calculations [3-7] at 7 and 8 TeV LHC energy together with 
exact NLO \cite{NLO} and NNLO \cite{NNLO} results. All results are 
with the same choice of parameters and including  
theoretical uncertainties from scale variation and other sources as described
in each paper [3-7]; these uncertainties however do not include uncertainties 
from parton distribution functions (pdf) or $\alpha_s$ which are 
extraneous to the theoretical method used 
and should be the same for all approaches. 
The results use the same value of top quark mass and the same MSTW 2008 
NNLO pdf \cite{MSTW} and $\alpha_s$ as implemented in LHAPDF. 
We note that the values chosen for some of these parameters were 
slightly different among the published results of Refs. [2-7] so we here  
have chosen common values for a better comparison. 
Note, however, that there is a very small change (at the per mille level) 
if one uses the published results with the different parameters, and thus no 
noticeable change in the comparison and no change in the conclusions reached.
Fig. \ref{theorytev14lhcplot} plots corresponding results at 14 TeV LHC energy 
(left) and 1.96 TeV Tevatron energy (right). 

The result in \cite{NK} is from a double-differential formalism and uses pQCD.
The result in \cite{ALLMUW} is from a method for the total cross section only 
and uses pQCD as does \cite{CCMMN}.
The results in \cite{AFNPY} and \cite{BFKS} use SCET for the 
double-differential and total cross section, respectively.
There is a fairly wide spread in the numbers and hence the degree of success 
of the various approaches.

The  result of Ref. \cite{NK} is very close to the exact NNLO;  
both the central values and the scale uncertainty are nearly the same.   
This is true for all collider energies, as can be seen from Figs. 1 and 2, 
and also for all top quark mass values from 130 GeV 
to 210 GeV as was checked with the results in Table III of \cite{NNLO}.
This is in addition to the known excellent agreement between exact and 
approximate results at NLO.
There is around 1\% or less difference between approximate and exact cross 
sections at both NLO and NNLO.

The excellent agreement of \cite{NK} with preliminary exact NNLO results 
for the Tevatron energy was 
already discussed in detail in \cite{NK12,NKckm12}. The additional agreement 
with the exact NNLO results for all LHC energies proves the validity of the 
method of Ref. \cite{NK} in general and the theoretical arguments in its 
support (see \cite{NK,NK12,NKckm12,NKRV1,NKRV2})   
and it shows that they are not restricted to a given energy of 
one collider. It is also well known that the results of \cite{NK} are in 
excellent agreement with cross section data from CDF and D0 at the Tevatron 
and from ATLAS and CMS at the LHC, see the figures in 
Refs. \cite{NK12,NKckm12} for several comparisons. 

The fact that the results of Ref. \cite{NK} are very close to the exact 
NNLO \cite{NNLO} was expected from various theoretical reasons that were 
discussed in detail in \cite{NK,NK12,NKckm12}. This agreement   
was expected from the study of the NLO approximation, from the comparison 
of 1PI and PIM results in \cite{NKRV1}, and from other arguments regarding the 
analytical structure of the results and their implementation
(see also discussion in \cite{NK,NK12,NKckm12,NKRV2}). 
A double-differential calculation for partonic (as contrasted to absolute) 
threshold as used in \cite{NK} has a lot of 
theoretical/analytical information (also useful for deriving distributions), 
generality, and potential for numerical accuracy.
This is an important point with clear consequences.   
Now that NNLO is fully known numerically \cite{NNLO} (though not analytically)
for the total cross section, 
the next step is to add the approximate N$^3$LO corrections 
(see \cite{NNNLO} for previous results). For differential calculations 
approximate NNLO is still the state-of-the-art and is likely to be 
practically indistinguishable from any future exact NNLO, 
but one can add N$^3$LO corrections to the differential distributions as well.

The stability of the theoretical NNLO approximate results in our 
formalism over the past decade \cite{NKRV1,NKRV2,NK,NKy,NK12} is 
notable; it is in contrast to the resummation formalism with the 
minimal presciption used in \cite{NNLO,CCMMN} which has produced 
widely-ranging results for the threshold corrections in the past.
The reliability and stability of the results from our formalism \cite{NK}
and near-identical value to exact NNLO is very important for several reasons:

(1) It provides confidence of application to other processes, in
particular single-top \cite{NKst}.

(2) The results have been used widely as backgrounds for many analyses 
(Higgs, etc).

(3) It means that we presently have near-exact NNLO $p_T$ and rapidity 
distributions.

Regarding point (1), the success of the formalism for single-top production 
in all three channels \cite{NKst,NK12,NKckm12} in describing the Tevatron and 
LHC data complements the confidence gained from the above comparison that 
approximate NNLO \cite{NKst} should also be a good approximation to exact NNLO 
for single top production. 
Regarding point (2), since the approximate NNLO results for both top-pair 
and single-top have been used as backgrounds in many Tevatron 
and LHC analyses, it is reassuring to know that any difference from exact NNLO 
is negligible and would not have materially affected these analyses. 
Finally, point (3) is also very important, and the remarkable success  
of the approximate NNLO distributions in describing Tevatron and LHC data 
further reenforces the theoretical arguments. In the next section, we discuss 
the top quark distributions and compare them with recent LHC data. 

\mysection{Top quark differential distributions}

\begin{figure}
\begin{center}
\includegraphics[width=8cm]{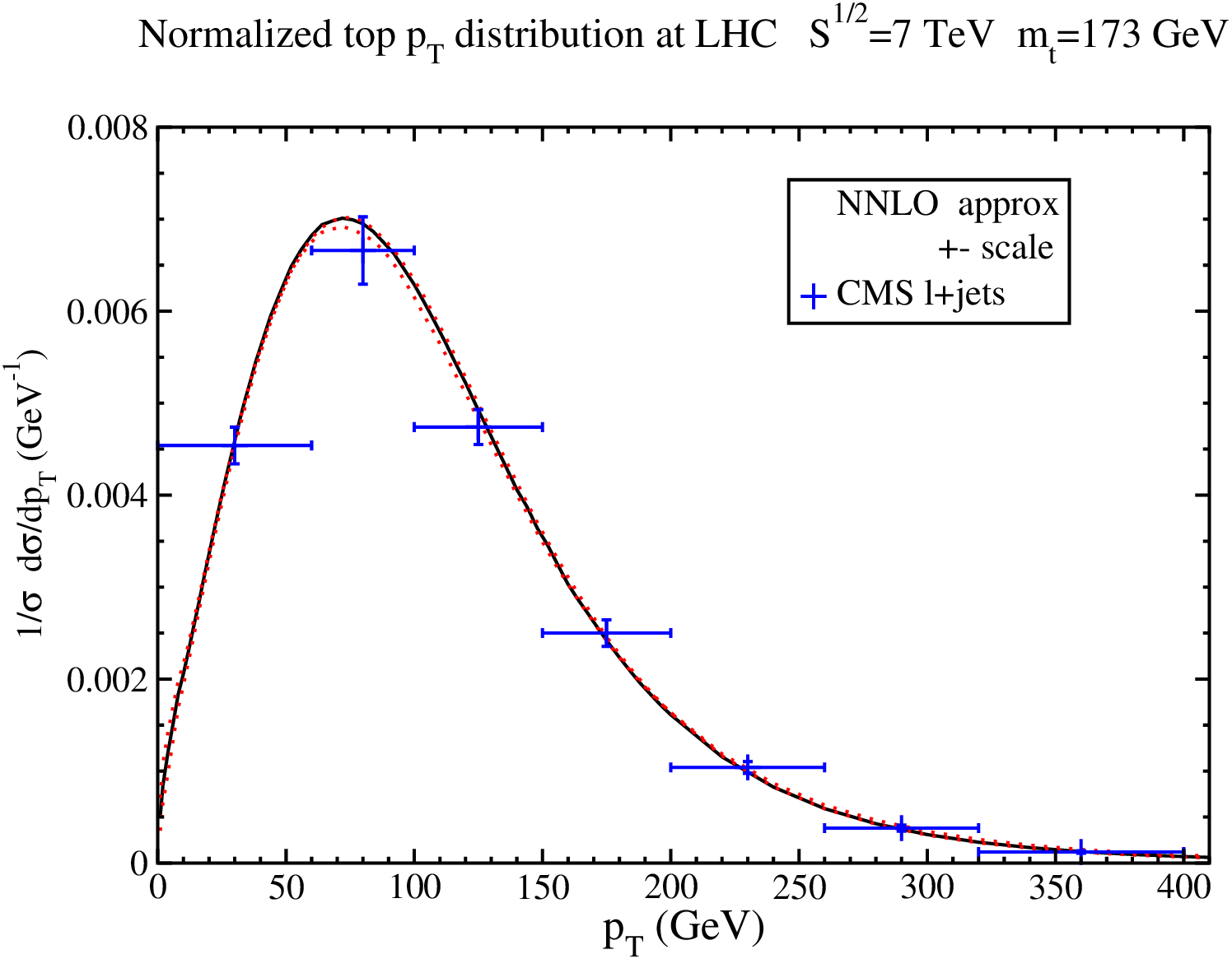}
\hspace{3mm}
\includegraphics[width=8cm]{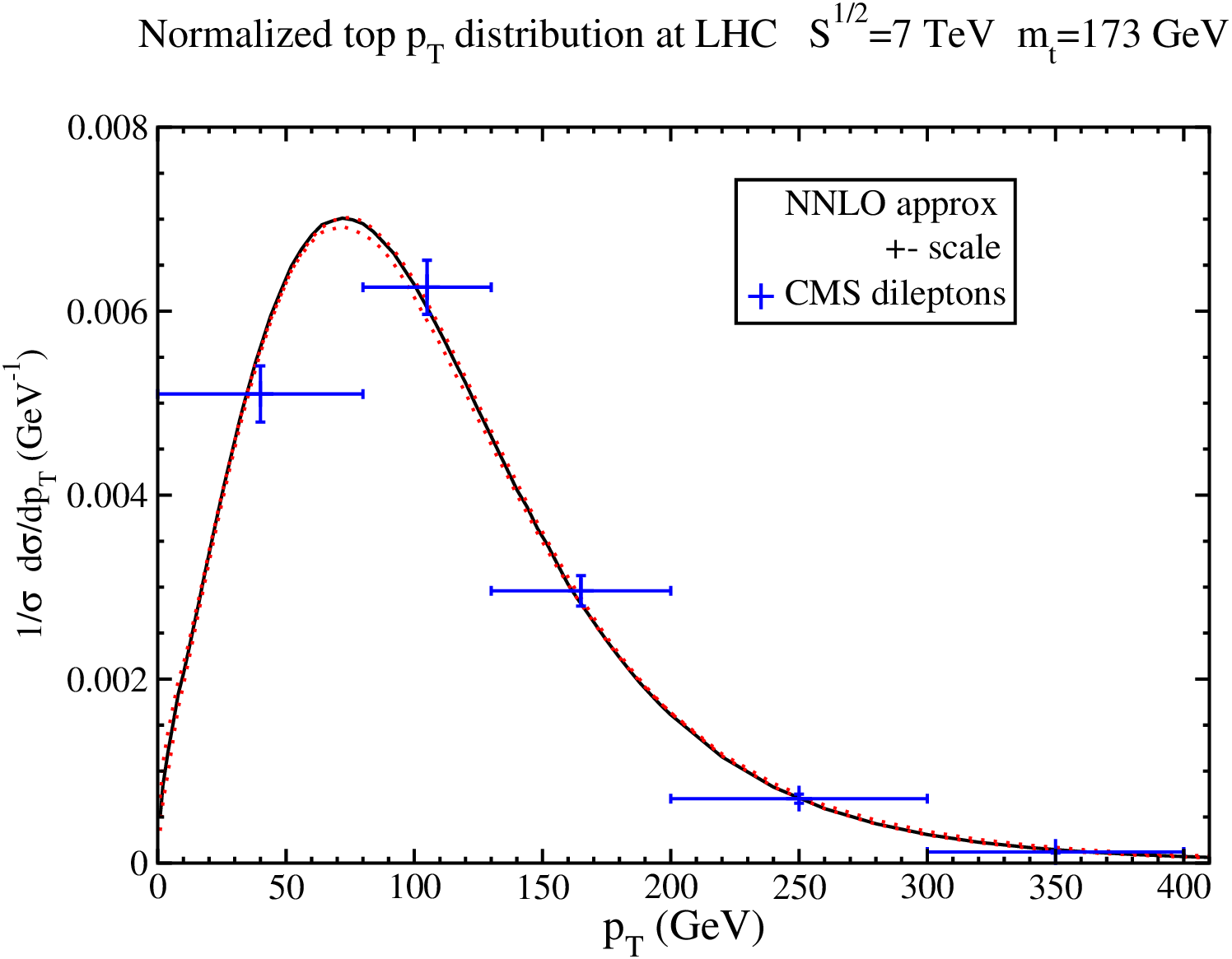}
\caption{Normalized approximate NNLO top-quark  $p_T$ distributions \cite{NK}
at the LHC and comparison with CMS data \cite{CMS} in the $\ell$+jets (left) 
and dilepton (right) channels.}
\label{pt7lhcplot}
\end{center}
\end{figure}

The soft-gluon approximation of \cite{NK,NKy} works very well both for 
total cross sections and differential distributions. 
The approximation is known to be excellent at NLO, with $\sim$1\% difference 
between NLO approximate and exact differential distributions,
see Fig. 2 in \cite{NK12}. Given the success of the approximation at NNLO for 
the total cross section, it is clear that the distributions should work very 
well at NNLO as well.

In Fig. \ref{pt7lhcplot} the theoretical top quark normalized transverse 
momentum distribution at approximate NNLO \cite{NK} for 7 TeV LHC energy 
is plotted and compared with recent data from the CMS collaboration \cite{CMS} 
in the $\ell$+jets and 
dilepton channels. The central result is for $\mu=m_t$ and the theoretical 
uncertainty from scale variation $m_t/2 < \mu < 2m_t$ is also 
displayed. The agreement of the LHC data in both CMS channels 
with the theoretical prediction is very good. 
The theoretical uncertainties are much smaller than the experimental error bars.
Similar results have also appeared for Tevatron energies and the 
agreement with D0 data \cite{D0} 
is excellent, see Fig. 4 in Ref. \cite{NKckm12}. 

\begin{figure}
\begin{center}
\includegraphics[width=8cm]{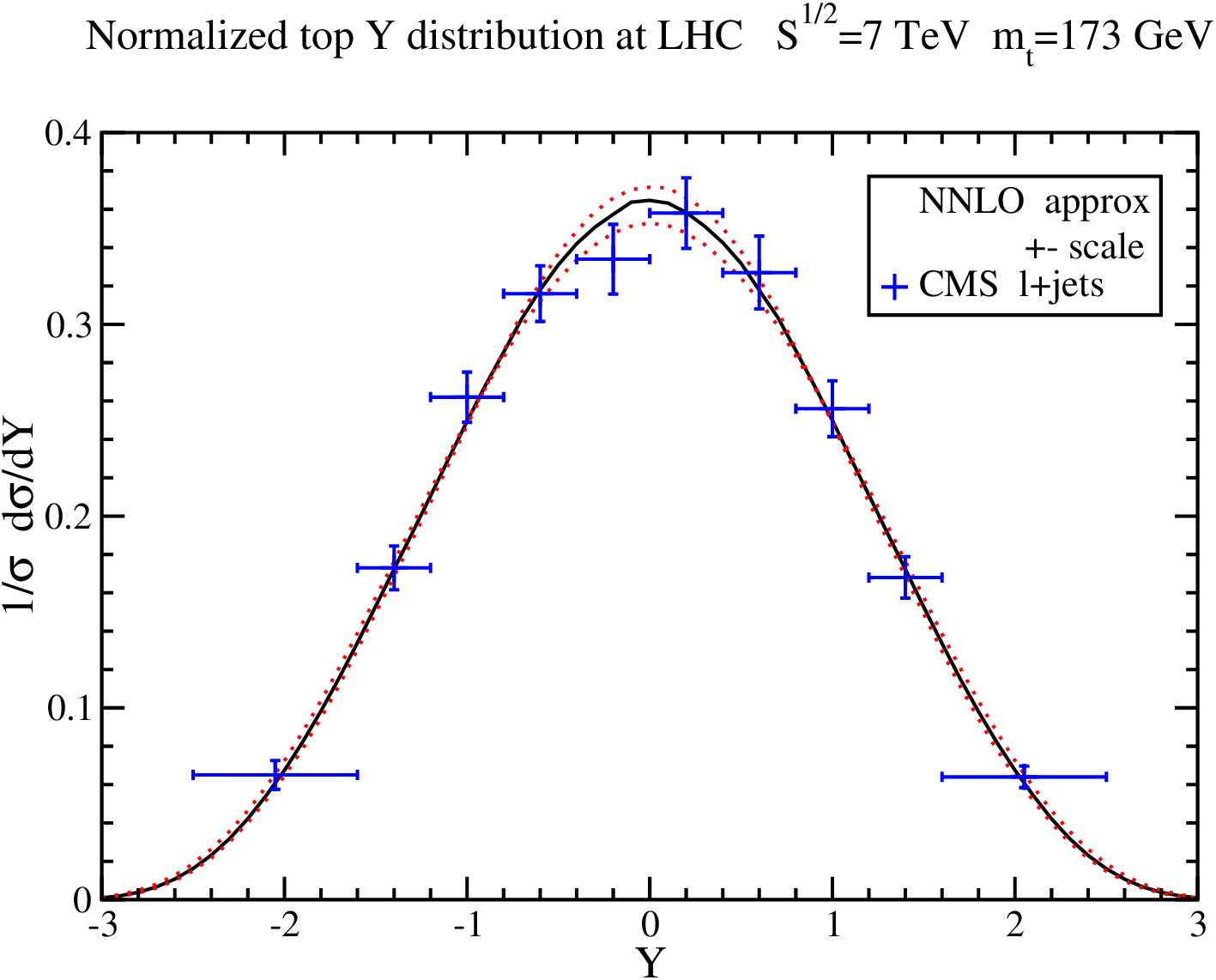}
\hspace{3mm}
\includegraphics[width=8cm]{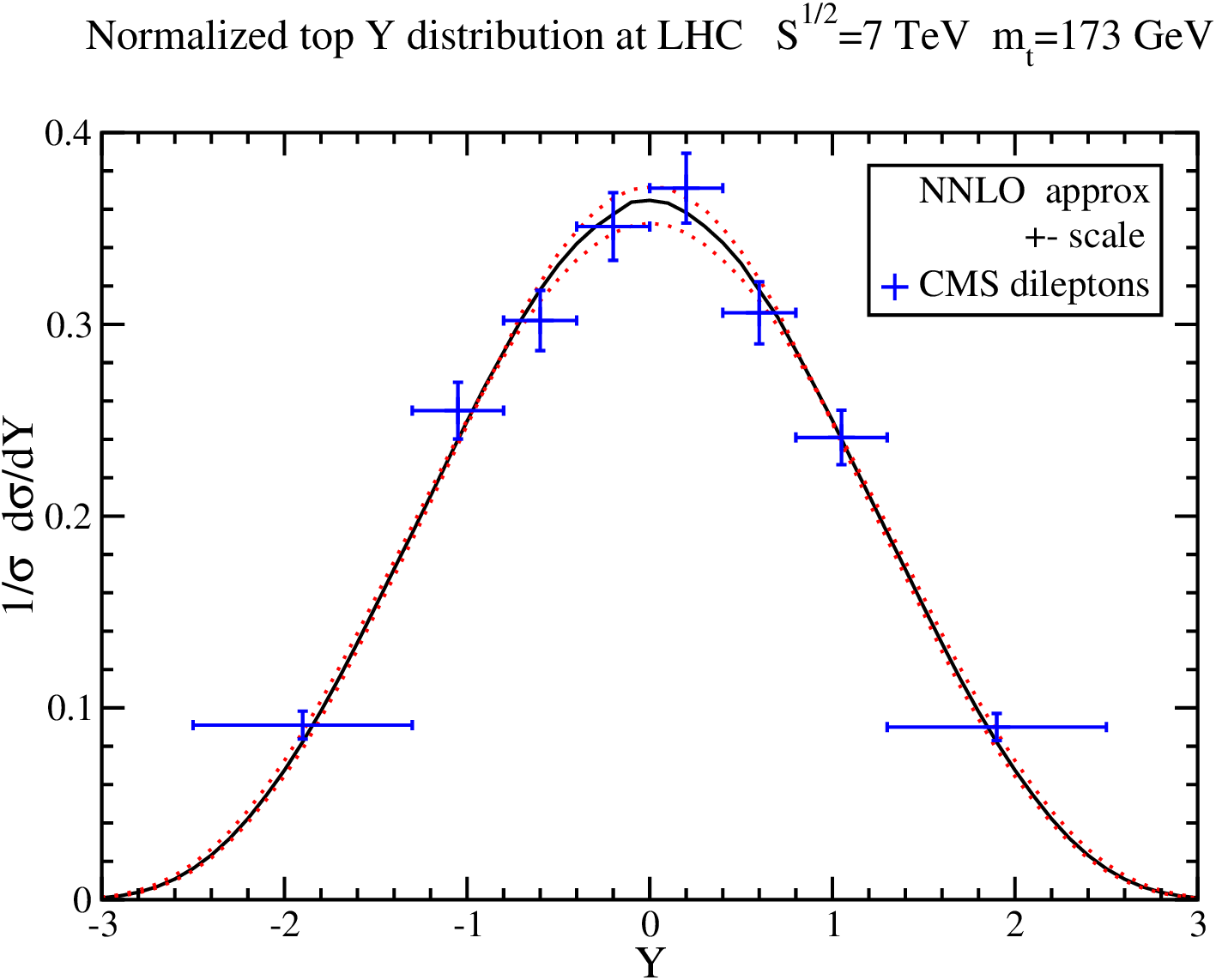}
\caption{Normalized approximate NNLO top-quark rapidity distributions 
\cite{NKy} at the LHC and comparison with CMS data \cite{CMS} 
in the $\ell$+jets (left) and dilepton (right) channels.}
\label{y7lhcplot}
\end{center}
\end{figure}

In Fig. \ref{y7lhcplot} the theoretical top quark normalized rapidity 
distribution, again with uncertainty from scale variation, at approximate 
NNLO \cite{NKy} for 7 TeV LHC energy is plotted 
and compared with recent data from the CMS collaboration in the $\ell$+jets and 
dilepton channels \cite{CMS}. Again, the agreement between LHC data and 
theoretical results is very good and the theoretical uncertainty is very small.

\mysection{Conclusions}

Various methods for soft-gluon resummation and approximate NNLO calculations 
for top-quark pair production have been discussed and compared.
We have shown in this paper that the soft-gluon approximation method 
of Ref. \cite{NK} works extremely well in approximating the exact NLO 
and NNLO total cross sections. The fact that the approximation also works 
extremely well for differential distributions at NLO provides confidence 
that the approximate NNLO distributions should be nearly indistinguishable from 
any future exact results.
The theoretical top-quark transverse momentum \cite{NK} and rapidity \cite{NKy} 
distributions have predicted very well the recent LHC data; the 
agreement is excellent. 
Future approximate N$^3$LO calculations will likely provide small additional 
enhancements for both total and differential cross sections and are
currently under study. 

\mysection*{Acknowledgements}
This material is based upon work supported by the National Science Foundation 
under Grant No. PHY 1212472.

\end{document}